\documentclass[
prl,twocolumn,preprintnumbers,
nobibnotes,nofootinbib,floatfix,
amsmath,amssymb,
longbibliography,superscriptaddress
]{revtex4-2}

\usepackage[utf8]{inputenc}
\usepackage{siunitx}
\usepackage{amsfonts}
\usepackage{braket}
\usepackage{mathtools}
\usepackage{cancel}
\usepackage{slashed}
\usepackage{pifont}
\usepackage{soul}
\usepackage{comment}
\usepackage{bbm}
\usepackage{bm}
\usepackage{macros_AS}
\usepackage[caption=false]{subfig}

\usepackage{yhmath} 

\usepackage{booktabs,array}
\usepackage{hhline}
\usepackage{siunitx}
\usepackage{graphicx}
\usepackage[caption=false]{subfig}
\usepackage{multirow}

\usepackage{dcolumn}
\usepackage{mathtools}
\usepackage{upgreek}
\newcolumntype{T}{D{.}{.}{10}}
\newcolumntype{E}{D{.}{.}{11}}
\newcolumntype{F}{D{.}{.}{5}}

\usepackage[normalem]{ulem}

\usepackage[dvipsnames]{xcolor}
\usepackage{hyperref}
\hypersetup{
    colorlinks=true,     
    linkcolor=blue,      
    citecolor=blue,      
    filecolor=blue,      
    urlcolor=blue        
}

\makeatletter 
\renewcommand\onecolumngrid{
\do@columngrid{one}{\@ne}%
\def\set@footnotewidth{\onecolumngrid}
\def\footnoterule{\kern-6pt\hrule width 1.5in\kern6pt}%
}
\renewcommand\twocolumngrid{
        \def\footnoterule{
        \dimen@\skip\footins\divide\dimen@\thr@@
        \kern-\dimen@\hrule width.5in\kern\dimen@}
        \do@columngrid{mlt}{\tw@}
}%
\makeatother

\begin{document}
\title{Gradient flow for parton distribution functions: first application to the pion}
\author{Anthony~Francis}
\affiliation{Institute of Physics, National Yang Ming Chiao Tung University, 30010 Hsinchu, Taiwan}
\author{Patrick~Fritzsch}
\affiliation{School of Mathematics, Trinity College, College Green, Dublin, Ireland}
\author{Robert~V.~Harlander}
\affiliation{Institute for Theoretical Particle Physics and Cosmology, TTK, RWTH Aachen University, Sommerfeldstr. 16, Aachen, 52074, Germany}
\author{Rohith~Karur}
\affiliation{Department of Physics, University of California, Berkeley, CA 94720, U.S.A}
\affiliation{Nuclear Science Division, Lawrence Berkeley National Laboratory, Berkeley, CA 94720, USA}
\author{Jangho~Kim}
\affiliation{Lattice Gauge Theory Research Center, Department of Physics and Astronomy, Seoul National University, Seoul 08826, Korea}
\author{Jonas~T.~Kohnen}
\affiliation{Institute for Theoretical Particle Physics and Cosmology, TTK, RWTH Aachen University, Sommerfeldstr. 16, Aachen, 52074, Germany}
\author{Giovanni~Pederiva}
\affiliation{J\"ulich Supercomputing Center, Forschungszentrum Jülich, Wilhelm-Johnen-Straße, 54245 Jülich, Germany}
\affiliation{Center for Advanced Simulation and Analytics (CASA), Forschungszentrum Jülich, Wilhelm-Johnen-Straße, 54245 Jülich, Germany}
\author{Dimitra~A.~Pefkou}
\email{dpefkou@berkeley.edu}
\affiliation{Department of Physics, University of California, Berkeley, CA 94720, U.S.A}
\affiliation{Nuclear Science Division, Lawrence Berkeley National Laboratory, Berkeley, CA 94720, USA}
\author{Antonio~Rago}
\affiliation{$\hbar$QTC \& IMADA, University of Southern Denmark, Campusvej 55, 5230 Odense M, Denmark}
\author{Andrea~Shindler}
\email{shindler@physik.rwth-aachen.de}
\affiliation{Institute for Theoretical Particle Physics and Cosmology, TTK, RWTH Aachen University, Sommerfeldstr. 16, Aachen, 52074, Germany}
\affiliation{Nuclear Science Division, Lawrence Berkeley National Laboratory, Berkeley, CA 94720, USA}
\affiliation{Department of Physics, University of California, Berkeley, CA 94720, U.S.A}
\author{Andr\'e~Walker-Loud}
\affiliation{Nuclear Science Division, Lawrence Berkeley National Laboratory, Berkeley, CA 94720, USA}
\affiliation{Department of Physics, University of California, Berkeley, CA 94720, U.S.A}
\author{Savvas~Zafeiropoulos}
\affiliation{Aix Marseille Univ, Universit\'e de Toulon, CNRS, CPT, Marseille, France}

\begin{abstract}
Parton distribution functions (PDFs) are central to precision QCD phenomenology. Their Mellin moments
can be computed on the lattice, but direct determinations using local operators, besides $\langle x \rangle$, face severe challenges from reduced hypercubic symmetry, limiting results to the lowest moments.  A recently proposed method resolves these issues using gradient flow.  We demonstrate the efficacy of this method by computing ratios of flavor non-singlet pion PDF moments up to $\braket{x^5}$, on four lattice spacings at $m_\pi \simeq 411$ MeV.
The moments and reconstructed PDF agree quantitatively with recent phenomenological extractions.
\end{abstract}

\preprint{TTK-25-21}

\maketitle

\textit{Introduction}: Parton distribution functions (PDFs) have a long history among hadron structure observables, dating back to the discovery of scaling in deep-inelastic scattering at SLAC in the late 1960s. Today, proton PDFs represent one of the dominant sources of uncertainty at several collider 
experiments~\cite{Amoroso:2022eow},
underscoring the urgency of improved theoretical input. Moreover, the PDFs of the pion remain much less constrained by experimental data~\cite{Conway:1989fs,NA10:1985ibr,Owens:1984zj,AURENCHE1989517,Sutton:1991ay,Gluck:1991ey,Wijesooriya:2005ir,Novikov:2020snp,H1:2010hym,ZEUS:2002gig,Barry:2018ort,Cao:2021aci,Barry:2021osv,Aicher:2010cb,Westmark:2017uig,Kotz:2023pbu,Kotz:2025lio}, since pions cannot be used as fixed targets. Their large-$x$ behavior, often parameterized as $(1-x)^{\beta}$, has been the subject of long-standing debate and is a strong motivation for upcoming measurements at Jefferson Lab~\cite{Montgomery:2017yua} and the Electron-Ion Collider~\cite{AbdulKhalek:2021gbh,Arrington:2021biu}. Model calculations typically predict values in the range $\beta=1\text{--}2$~\cite{Ezawa:1974wm,Landshoff:1973pw,Gunion:1973ex,Farrar:1979aw,Berger:1979du,Shigetani:1993dx,Szczepaniak:1993uq,Davidson:1994uv,Hecht:2000xa,Melnitchouk:2002gh,Noguera:2015iia,Hutauruk:2016sug,Hobbs:2017xtq,deTeramond:2018ecg,Bednar:2018mtf,Lan:2019vui,Lan:2019rba,Chang:2020kjj,Cui:2020tdf,Kock:2020frx,Cui:2022bxn,Albino:2022gzs,Ahmady:2022dfv,Pasquini:2023aaf,Lu:2023yna}.

The only systematically improvable first-principles approach to calculate hadron structure functions is lattice QCD. In Euclidean space, where lattice QCD is formulated, the non-local operators defining PDFs collapse to a point, making their direct computation impossible. 
A natural approach is to consider Mellin moments,
\be 
\langle x^{n-1} \rangle_q(\mu) = \int dx \, x^{n-1} \, q(x,\mu)\,,
\label{eq:mellin_def}
\ee
with $q(x,\mu)$ a PDF at renormalization scale $\mu$.
They can be obtained from matrix elements of local twist-2 operators,
\be
O^{rs}_n(x) \equiv i^{n-1} \bar{\psi}^r \gamma_{\{\mu_1} \overleftrightarrow{D}_{\mu_2}\ldots\overleftrightarrow{D}_{\mu_n\}} \psi^s(x)\,,
\label{eq:twist2}
\ee
with $\psi^r$ a quark field of flavor $r$, $\overline{\psi}^s$ an antiquark field of flavor $s$, $\overleftrightarrow{D}_{\mu} = \tfrac{1}{2}(D_\mu - \overleftarrow{D}_\mu)$ the forward-backward covariant derivative, and $\{\cdots\}$ denoting normalized symmetrization over Lorentz indices.  

In the flavor non-singlet case there is no mixing with gluonic twist-2 operators, and the moments are obtained from reduced matrix elements of the trace-subtracted operator of Eq.~\eqref{eq:twist2} between the hadronic states of interest.

\begin{table*}[t]
\begin{center}
\begin{tabular}{c|ccccccc}
\hline
\hline
  label & $a~(\text{fm})$  & $t_0/a^2$ & $\beta$ & $\kappa_{ud}=\kappa_s$ & $T/a \times \left(L/a\right)^3$  & N$_{\text{cfg}}$ & $\tau_s/a$ \\
 \hline
  {\tt a12m412\_mL6.0}  & 0.12 & 1.4868(04) & 3.685& 0.1394305 & $96\times 24^3$ & 838 & 35,40\\
  {\tt a094m412\_mL6.2} & 0.094 & 2.4400(01)  & 3.8  & 0.138963 & $96\times 32^3$ & 700 & 35,40\\
  {\tt a077m412\_mL7.7} & 0.077 & 3.6239(12) & 3.9 & 0.138603 & $96\times 48^3$ & 400 & 40,42\\
  {\tt a064m412\_mL6.4} & 0.064 & 5.2471(26) & 4.0  & 0.138272 & $96\times 48^3$ & 500 & 40,42\\
  \hline
\hline

\end{tabular}
\caption{Details of the SWF lattice ensembles used in this work, which were generated by the OpenLat initiative~\cite{Francis:2022hyr,Cuteri:2022oms,Cuteri:2022erk,Francis:2023gcm}, including the total number of gauge configurations, N$_{\text{cfg}}$, and the specific source-sink separations, $\tau_s/a$, on which the three-point functions were measured.}
\label{tab:gauges} 
\end{center}
\end{table*}

The computation of Mellin moments from local operators has been an active area of lattice QCD research for several decades starting in the 1980s~\cite{Kronfeld:1984zv,Martinelli:1987zd,Martinelli:1988rr}. 
As these early studies already pointed out, additional challenges arise because the O($4$) symmetry of the continuum is reduced to the hypercubic symmetry H($4$) on the Euclidean lattice.  

Accessing the $x$-dependence of PDFs with lattice QCD has seen significant progress in recent decades, with numerous new approaches proposed, including methods to compute Mellin moments from derivatives of Ioffe-time distributions~\cite{Aglietti:1998mz,Detmold:2005gg,Ji:2013dva,Monahan:2015lha,Radyushkin:2016hsy,Orginos:2017kos,Chambers:2017dov,Braun:2007wv,Ma:2014jla,Liang:2019frk,Izubuchi:2018srq,Karpie:2018zaz,Joo:2019jct,Joo:2019bzr,Joo:2020spy,Fan:2020nzz,HadStruc:2021qdf,Gao:2022iex,Gao:2022uhg,HadStruc:2024rix}.

In this work we investigate a recently proposed method~\cite{Shindler:2023xpd} that employs the gradient flow as an intermediate regulator to compute Mellin moments, and their ratios, of arbitrary order in lattice QCD.  
We present the first numerical demonstration of this approach for the flavor non-singlet pion PDF, thus establishing its practical applicability and effectiveness in lattice calculations.   
With a modest computational cost, 
we achieve competitive uncertainties for the lowest moments, extend the direct computation of Mellin moments from local operators up to $n=6$, and show that a simple PDF reconstruction from these ratios yields results consistent with phenomenological extractions. Our study highlights the potential of this approach to extend lattice determinations of PDFs beyond the current state of the art. In the future, these moments can also be incorporated directly into global PDF fits alongside experimental and/or lattice QCD data, e.g. Ref.~\cite{JeffersonLabAngularMomentumJAM:2022aix}.

\textit{Flowed moments}: 
The reduction of O($4$) to H($4$) symmetry on the Euclidean lattice manifests itself in the form of power-divergent operator mixing in calculations of PDF moments.  
To address this problem, Ref.~\cite{Shindler:2023xpd} proposed to use the gradient flow (GF) for the gauge~\cite{Narayanan:2006rf,Luscher:2010iy,Luscher:2011bx} and fermion~\cite{Luscher:2013cpa} fields. 
The flow time $t$ acts as an intermediate regulator of ultraviolet divergences: flowed operator matrix elements are finite in the continuum limit if the operator is purely gluonic~\cite{Luscher:2010iy}, and renormalize multiplicatively if fermion fields are present.  
In the latter case, the renormalization is particularly simple, depending only on the fermion content of the operator~\cite{Luscher:2013cpa}. 
Once the fermion fields are renormalized, one can safely take the continuum limit $a \to 0$ of matrix elements of the flowed operator $\mcO_n(t,x)$, defined as in Eq.~\eqref{eq:twist2} with flowed fields.  

In the continuum, these flowed matrix elements can then be related to their physical $t=0$ counterparts through the short flow-time expansion (SFTX)~\cite{Suzuki:2013gza,Luscher:2013vga} 
and the corresponding matching coefficients.  
The matching is simplified and purely multiplicative, making use of the restored $O(4)$ symmetry and its irreducible representations (irreps), by computing matrix elements of flowed traceless twist-2 operators with symmetrization over Lorentz indices, $\widehat{\mcO}_n^{rs}(t,x)$.

Multiplicative renormalization of the flowed fermion fields can be avoided by considering ratios of matrix elements with identical fermion content.  
In this way, ratios of PDF moments, e.g. in the $\overline{\MS}$ scheme, can be obtained as
\bea 
\frac{\braket{x^{n-1}}^{\overline{\MS}}(\mu)}{\braket{x^{m-1}}^{\overline{\MS}}(\mu)} 
= \frac{\zeta_{m}(t,\mu)}{\zeta_{n}(t,\mu)} \,
  \frac{\braket{x^{n-1}}(t)}{\braket{x^{m-1}}(t)} + O(t)\,,
  \label{eq:momentratio}
\eea
where the matching coefficients $\zeta_n(t,\mu)$ were calculated at next-to-leading order (NLO) in Ref.~\cite{Shindler:2023xpd},\footnote{In Ref.~\cite{Shindler:2023xpd} the matching coefficients $\zeta_n$ are denoted by $c_n$.}  
and the $O(t)$ terms denote contributions from higher-dimensional operators in the SFTX. 

The ratio of flowed moments is directly related to the ratio of reduced matrix elements of the flowed operator, analogous to the $t=0$ case. It is free from ultraviolet divergences and can be obtained in the continuum limit by keeping the flow time $t$ fixed in physical units. A discussion of the continuum limit and cutoff effects is deferred to the End Matter~\ref{sec:EM}. 

In this work we perform the matching at NNLO up to $n=6$. The coefficients
$\zeta_{n}(t,\mu)$ of Eq.~\eqref{eq:momentratio} are determined perturbatively
using the method of projectors~\cite{Gorishnii:1983su,Gorishnii:1986gn}, first
applied in the context of the gradient flow in
Ref.~\cite{Harlander:2018zpi}. It leads to scalar two-loop momentum integrals
which only depend on the flow time $t$ and the space-time parameter
$D=4-2\epsilon$; they can be solved analytically using standard tools (see,
e.g., Ref.~\cite{Artz:2019bpr}). Requiring finite $\overline{\MS}$ renormalized
matching coefficients confirms the well-known two-loop anomalous dimensions of
the twist-2 operators~\cite{Floratos:1977au,Gonzalez-Arroyo:1979guc}.  At NLO,
we find agreement with the matching coefficients of
Refs.~\cite{Suzuki:2013gza,Makino:2014taa,Shindler:2023xpd}. At NNLO, the
results have been known only for the first two moments
($n=1,2$)~\cite{Harlander:2018zpi,Borgulat:2023xml};
they also agree with our findings. As another
check, we verified that our final results for the renormalized matching
coefficients are independent of the QCD gauge parameter. The analytical
results for the matching coefficients as well as more details of their
calculation will be deferred to a forthcoming publication.

The strategy to follow can thus be summarized as: (1) determine the ratios of flowed moments that appear on the RHS of Eq.~\eqref{eq:momentratio} from lattice QCD using the standard techniques described below, (2) take the continuum limit at fixed flow time $t$, (3) match to the $\overline{\MS}$ scheme at scale $\mu$ using the coefficients $\zeta_n(t,\mu)$ in Eq.~\eqref{eq:momentratio}, and (4) analyze the resulting ratio of PDF moments as a function of $t$; if a residual $t$-dependence remains, extrapolate to $t \to 0$ to obtain the physical result.
In this Letter we present the main results of this procedure, while additional details of each step will be provided in a forthcoming companion publication~\cite{PRD2025}.

\textit{Lattice QCD calculation}: The matching described in the previous section is independent of the choice of Lorentz indices of the flowed twist-2 operators. To optimize the signal-to-noise ratio, we compute matrix elements of operators with only temporal Euclidean components, $\widehat{O}_{n} = \widehat{O}_{n,4\ldots 4}$, made traceless by subtracting the appropriate spatial or mixed components~\cite{Shindler:2023xpd}.
For this operator choice, no external three-momentum is required, and the flowed moments are related to matrix elements via
\bea 
\frac{\braket{x^{n-1}}(t)}{\braket{x^{m-1}}(t)} 
= \frac{(-1)^{n-m}}{m_h^{\,n-m}}
  \frac{\braket{h({\bf 0})|\widehat{O}_{n,4\ldots4}(t)|h({\bf 0})}}
       {\braket{h({\bf 0})|\widehat{O}_{m,4\ldots4}(t)|h({\bf 0})}}\,,
\label{eq:momentratioflow}
\eea
where the factor $(-1)^{n-m}$ accounts for the Euclidean metric. 
To extract the matrix elements of Eq.~\eqref{eq:momentratioflow} for the pion we compute the connected three-point functions
\bea
    C_n^{3\text{-pt}}(& x_4 & = \tau_s,y_4 = \tau_{\mcO}; t) =  \nonumber \\
    & & a^6\sum_{\bx,\by} \braket{P^{du}(\bx,x_4)\widehat{O}_n(\by,y_4;t) P^{ud}(0) }_c\,,
    \label{eq:3ptfull} 
\eea
projected to zero spatial momentum for all $n$.  
Here $P^{ud}(x) = \overline{\psi}_u(x)\gamma_5\psi_d(x)$ denotes an interpolating operator with the quantum numbers of a $\pi^+$ state.  

On a hypercubic lattice, the O($4$) representation, when restricted to H($4$), 
decomposes into a direct sum of H($4$) irreps.  
This implies that different discretizations of traceless, Lorentz-symmetric operators are possible.  
Although all such choices share the same continuum limit, 
this freedom could in principle be exploited to reduce discretization effects.  
Here we restrict to the standard O($4$) irrep for all $n$, leaving a numerical study of alternative discretizations to future work.

We compute three-point functions for $n \in \{2,3,4,5,6\}$ using four Stabilized Wilson Fermion (SWF) ensembles generated by the OpenLat initiative~\cite{Cuteri:2022erk,Cuteri:2022oms,Francis:2022hyr,Francis:2023gcm}.  
These ensembles span four lattice spacings, $a \simeq [0.12, 0.094, 0.077, 0.064]$~fm, with the lattice scale set from $t_0/a^2$~\cite{Luscher:2010iy} and converted to physical units using $\sqrt{t_0} = 0.14474(57)$ fm~\cite{FlavourLatticeAveragingGroupFLAG:2024oxs,RBC:2014ntl,RQCD:2022xux,Bruno:2016plf,BMW:2012hcm}. 
They are generated with $N_f=3$ degenerate dynamical quarks, tuned such that the pion mass is $m_\pi \simeq 411~\text{MeV}$. 
The number of independent configurations and other run parameters for each ensemble are listed in Table~\ref{tab:gauges}. Further technical details of the lattice QCD calculation are given in the End Matter.
\begin{figure}[!]
\centering
\includegraphics[width=0.5\textwidth]{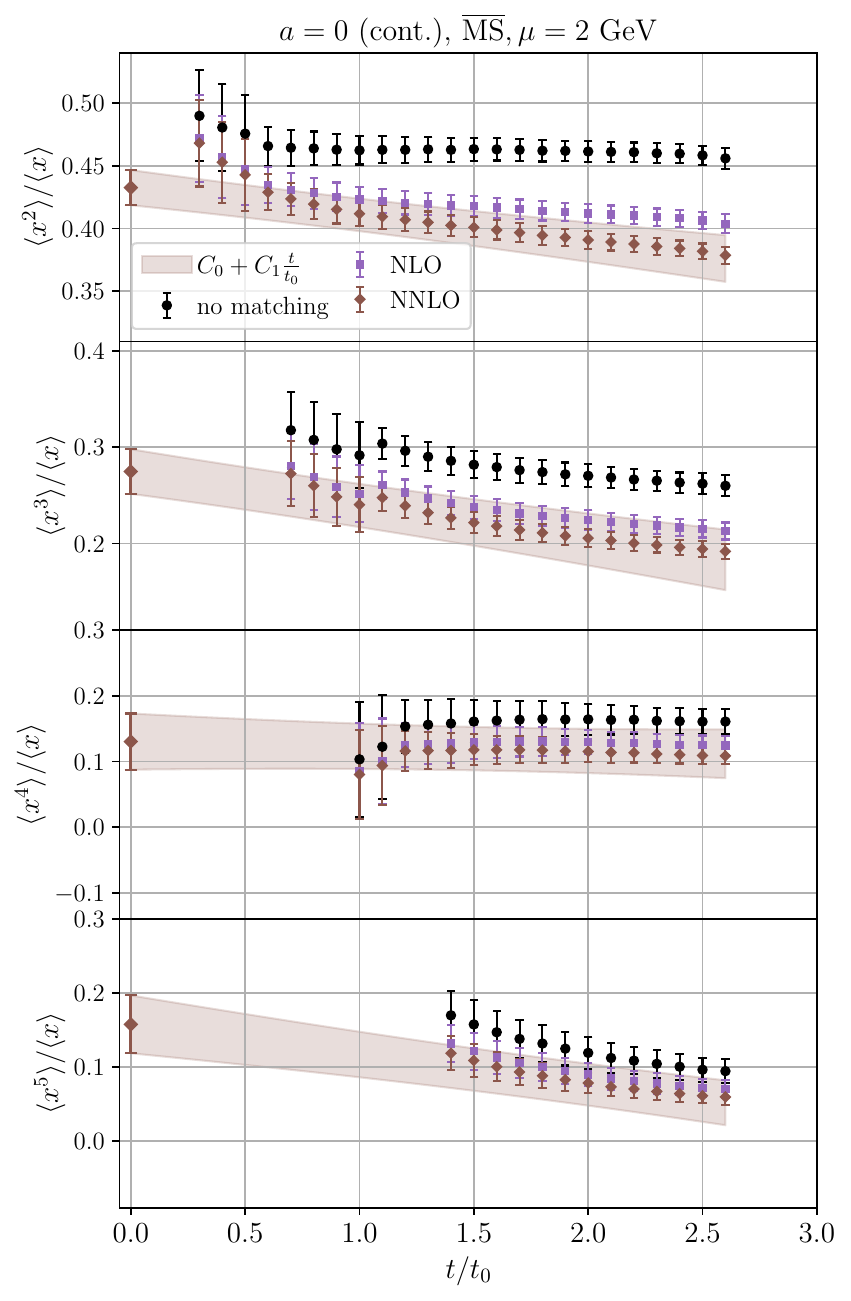}
\caption{Continuum-extrapolated ratios of Mellin moments $\langle x^{n-1}\rangle/\langle x\rangle$ for $n=3,4,5,6$ as functions of the flow time $t/t_0$.  
Black points show the continuum limit without matching, while purple and brown points correspond to NLO and NNLO matching, respectively, in the $\overline{\text{MS}}$ scheme at $\mu=2~\text{GeV}$.  
The shaded bands indicate the outcome of the fit procedure, where multiple fits of the form $C_0 + C_1\, t/t_0$ over different ranges of the NNLO matched ratios are combined with equal weights to estimate the physical $t \to 0$ limit.}
\label{fig:allfinal_match}
\end{figure}

We are interested in determining ratios of matrix elements for the ground state, $A_n(t)/A_2(t)$,\footnote{We set to $m=2$, as we do not find any advantage in other choices.} where $A_n(t)\equiv \left\langle \pi(\bzero) | \widehat{O}_n(t) | \pi(\bzero) \right\rangle$.
To isolate the ground state we assume the limit $\tau_s \gg \tau_{\mcO} \gg 0$, which ensures ground-state dominance. In this regime, for each flow time $t/a^2$, the ratio of three-point functions becomes
\bea
\frac{C_n^{3\text{-pt}}(\tau_s,\tau_{\mcO}; t)}{C_2^{3\text{-pt}}(\tau_s,\tau_{\mcO}; t)} \simeq
\frac{A_n(t)}{A_2(t)} + \cdots \,,
\label{eq:ratio_ground}
\eea
where the ellipsis denotes excited-state contributions.
We then look for plateau-like behavior as a function of the operator insertion time $\tau_{\mcO}$ for each source-sink separation $\tau_s$. 
As an effective estimator of $A_n(t)/A_2(t)$, we use an average over all $(\tau_s,\tau_{\mcO})$ values of the ratios with $\tau_s/2-2\leq \tau_{\mcO} \leq \tau_s/2+2$, with an additional systematic error equal to the half-difference between the maximum and minimum central value of the averaged together ratios.
Further details on this choice and on excited states contamination are given in the End Matter.

\begin{figure}
\includegraphics[width=\columnwidth]{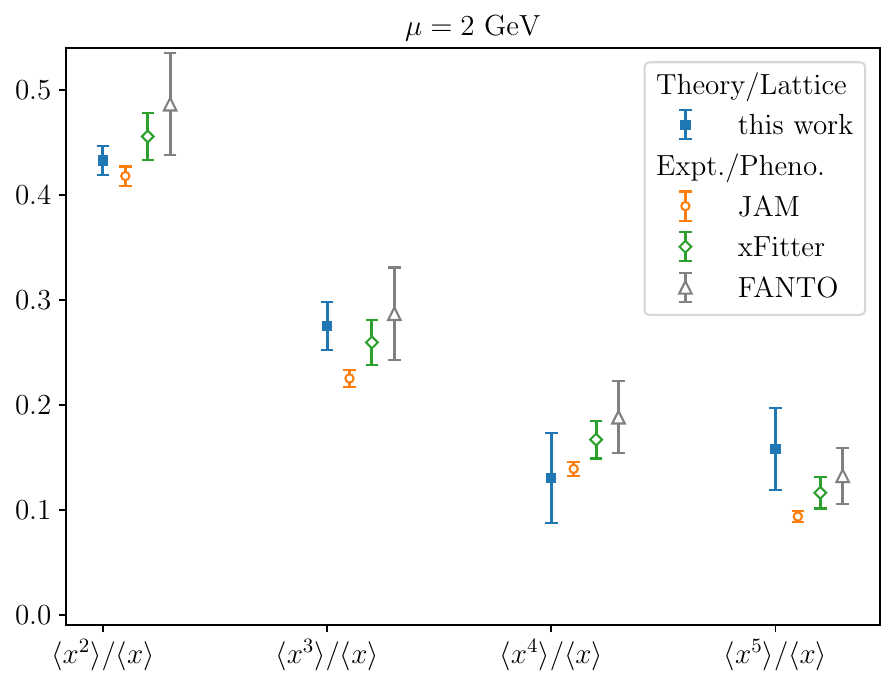}
\caption{Comparison of our continuum-extrapolated results from Table~\ref{tab:results} for the pion PDF moment ratios $\braket{x^{n-1}}/\braket{x}$ at $\mu=2~\text{GeV}$ (blue squares), obtained at a pion mass of $m_{\pi}\simeq 411~\text{MeV}$, with phenomenological extractions of the valence sector from JAM (orange circles)~\cite{Barry:2021osv}, xFitter (green diamonds)~\cite{Novikov:2020snp}, and FANTO (grey triangles)~\cite{Kotz:2025lio}, all obtained from global fits to experimental data.}
\label{fig:comparepheno}
\end{figure}

\textit{Moment results}: To obtain the flowed moment ratios we multiply our final results $A_n(t)/A_2(t)$ by the appropriate factors involving the pion mass in lattice units, as defined in Eq.~\eqref{eq:momentratioflow}.  
The resulting ratios are extrapolated to the continuum limit, assuming cutoff effects of $\mathcal{O}(a^2)$. The outcome is shown in Fig.~\ref{fig:allfinal} in the End Matter, where results for individual ensembles at each lattice spacing are displayed together with the extrapolated $a \to 0$ values (black circles). Further details on the continuum extrapolation and the choice of the minimal value $\left(t/t_0\right)_{\text{min}}$ are provided in the End Matter.

Finally, we multiply the continuum-extrapolated values of each ratio by the matching factors to the $\overline{\MS}$ scheme at the scale $\mu=2~\text{GeV}$, and examine the results for residual flow-time dependence.  
This is shown in Fig.~\ref{fig:allfinal_match}, where we display the continuum results before matching (black circles), and after matching at NLO (purple squares) and NNLO (brown diamonds).  
The NNLO-matched results are then extrapolated to $t=0$ using a linear fit in $t/t_0$.  
The fit ranges are taken as all possible intervals within $[\left(t/t_{0}\right)_{\mathrm{min}},2.6]$, where $\left(t/t_{0}\right)_{\mathrm{min}}$ for each $n$ are defined in Fig.~\ref{fig:allfinal}, with a minimum fit range in $t/t_0$ of at least $0.8$.  
Each fit with $p$-value $>0.1$ is weighted 
equally as a conservative choice, and the shaded bands represent the combined outcome of this procedure. 

The results of this analysis constitute our final estimates for $\braket{x^{\,n-1}}/\braket{x}$ in the $\overline{\text{MS}}$ scheme at $\mu=2~\text{GeV}$, and are listed in Table~\ref{tab:results}.

\begin{table}[h]
\begin{center}
\begin{tabular}{c|cccc}
\hline\hline
  $n$ & 3  & 4 & 5 & 6 \\
\hline
 $\braket{x^{n-1}}/\braket{x}$ 
   & 0.433(14) & 0.275(23) & 0.130(43) & 0.158(39) \\
\hline\hline
 & \multicolumn{2}{c}{Fit with $n_{\text{max}}=6$} & \multicolumn{2}{c}{Fit with $n_{\text{max}}=5$} \\
\hline
$\alpha$ & \multicolumn{2}{c}{-0.48(10)} & \multicolumn{2}{c}{-0.47(11)} \\
$\beta$  & \multicolumn{2}{c}{0.93(16)}   & \multicolumn{2}{c}{0.97(17)}   \\
\hline\hline
\end{tabular}
\caption{\label{tab:results} 
Top: final results for $\braket{x^{n-1}}/\braket{x}$ in the $\overline{\MS}$ scheme at $\mu=2~\text{GeV}$.  
Bottom: parameters of the PDF reconstruction fits (see Fig.~\ref{fig:pdfcomparepheno}), performed using the ansatz of Eq.~\eqref{eq:PDFparam}, either including ($n_{\text{max}}=6$) or excluding ($n_{\text{max}}=5$) the highest-order moment ratio.}
\end{center}
\end{table}

In Fig.~\ref{fig:comparepheno}, we compare our results with recent phenomenological determinations~\cite{Barry:2021osv,Novikov:2020snp,Kotz:2025lio}\footnote{In the case of Ref.~\cite{Barry:2021osv}, the comparison employs the NLO+NLL double Mellin method.}.  
Perfect agreement is not expected, since our calculation is performed at a heavier-than-physical pion mass of $m_\pi \simeq 411~\text{MeV}$.  
Nevertheless, guidance on the expected pion-mass dependence can be drawn from earlier lattice QCD studies for $n=2,3,4$, which, despite large statistical uncertainties, indicate only a mild dependence for the ratios considered here~\cite{Brommel:2005ee,Brommel:2007zz}, as discussed in the Supplemental Material (SM).  

\begin{table*}[t]
\begin{center}
\begin{tabular}{ccccccccc}
\hline\hline
 & $N_f$ & $m_{\pi}$~[MeV] & $a$~[fm]& $\braket{x^2}/\braket{x}$ &  $\braket{x^3}/\braket{x}$ &  $\braket{x^3}/\braket{x^2}$ &  $\braket{x^4}/\braket{x}$ &  $\braket{x^4}/\braket{x^2}$ 
 \\
\hline
this work & 3 & 411 & cont. & 0.433(14) & 0.275(23) & 0.636(50) & 0.130(43) & 0.301(98) \\
ETMC~\cite{Alexandrou:2020gxs} & 2+1+1 & 260 & 0.093 & 0.30(11) &&&& \\
ETMC~\cite{Alexandrou:2021mmi} & 2+1+1 & 260 & 0.093 &  & 0.092(71) & 0.22(17) && \\
HadStruc~\cite{Joo:2019bzr} & 2+1 & 415 & 0.127 & 0.370(47) & 0.224(16) &  &0.059(47)  & \\
BNL~\cite{Gao:2022iex} & 2+1 & 140& cont.\footnote{The continuum limit is estimated using calculations at $a=0.076$ fm with $m_\pi=140$ MeV, and 
at $a=0.04$ and $0.06$ fm with $m_\pi=300$ MeV.} & & & & & 0.351(34) 
\end{tabular}
\caption{Comparison with twist-2 operator determinations from ETMC~\cite{Alexandrou:2020gxs,Alexandrou:2021mmi}, derivatives of Ioffe time distributions from HadStruc~\cite{Joo:2019bzr} and BNL~\cite{Gao:2022iex}. For the results of Refs.~\cite{Joo:2019bzr}, covariances between the moments are not available and have been neglected in the shown ratios.
\label{tab:compare}
}
\end{center}
\end{table*}
A comparison with other direct lattice QCD calculations is presented in Table~\ref{tab:compare}.
Refs.~\cite{Alexandrou:2020gxs,Alexandrou:2021mmi} employ states boosted in multiple directions to avoid power-divergent mixing with lower-dimensional operators, which leads to signals that are significantly noisier than ours.  
Refs.~\cite{Gao:2022iex,Joo:2019bzr} instead use Ioffe-time distributions to extract PDF moments. 
The discrepancies between different lattice QCD computations are not worrisome because calculations are done with different pion masses, different number of dynamical flavors, and in some cases with systematics not completely under control.

While this work is a proof-of-principle calculation, we nevertheless identify potential sources of systematic effects to be investigated further once the statistical precision is increased and more data becomes available. In addition to the conservatively treated excited-state contamination,
in our calculation we perform a robust continuum extrapolation at an unphysical pion mass.  
Cutoff effects grow, as expected, at small flow times; however, in the region where they are under control we observe that the relative difference between NLO and NNLO corrections is about $5\%$, and that the $t \to 0$ extrapolation is stable with respect to variations of the fit range in flow time. This is reassuring for future calculations with higher precision.  

\begin{figure}
\includegraphics[width=\columnwidth]{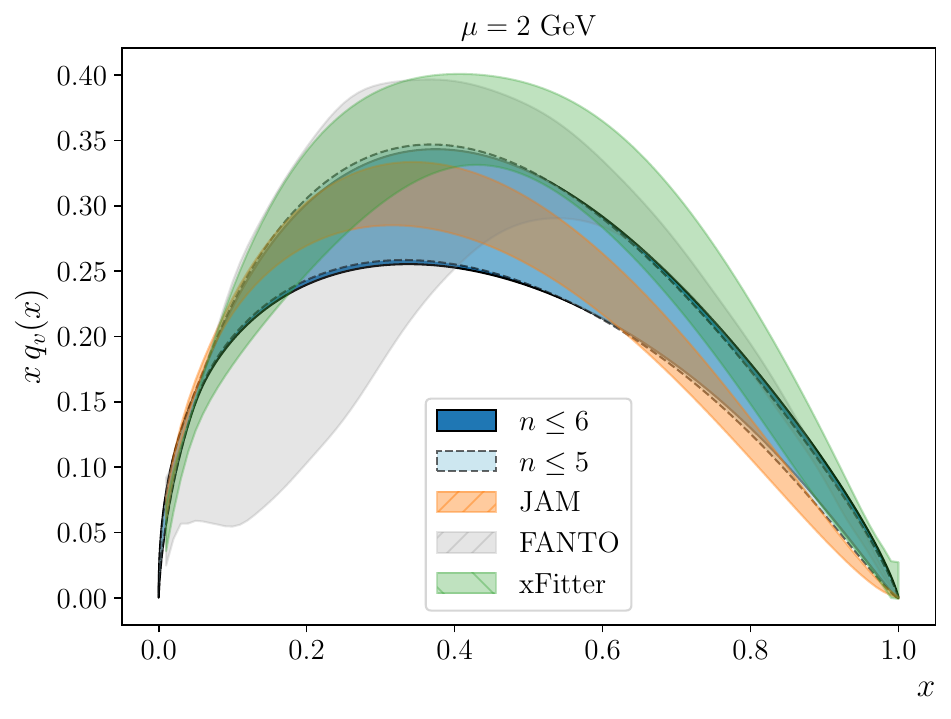}
\caption{Reconstructed pion valence PDF $q_v(x)\propto x^\alpha(1-x)^\beta$ obtained by fitting the parameters $\alpha$ and $\beta$ to the ratios of moments computed in this work.  
The blue band uses all available moments up to $n=6$, while the light-blue band excludes the highest ratio $\langle x^5\rangle/\langle x\rangle$.  
For comparison we also show phenomenological extractions from JAM, xFitter, and FANTO~\cite{Barry:2021osv,Novikov:2020snp,Kotz:2025lio}, respectively.}
\label{fig:pdfcomparepheno}
\end{figure}

\textit{PDF reconstruction}: Given our results for the ratios of moments in Table~\ref{tab:results}, we attempt a reconstruction of the PDF using the simple ansatz~\cite{Sutton:1991ay,Detmold:2003tm}
\be
q_v(x) \propto x^{\alpha}(1-x)^{\beta},
\label{eq:PDFparam}
\ee
and fit our data to the analytic dependence of the moments on $\alpha$ and $\beta$.  
As expected, the fitting procedure is non-trivial: while Gaussian priors help stabilize the fit, the results remain highly sensitive to the prior width of $\alpha$.  
We also find that excluding the highest-order moment ($n=6$) has little impact on the extracted values of $\alpha$ and $\beta$, shown in Table~\ref{tab:results}, indicating that higher moments affect these parameters only weakly unless determined with higher precision. 

The resulting PDF, shown in Fig.~\ref{fig:pdfcomparepheno}, is in reasonable agreement with phenomenological extractions~\cite{Barry:2021osv,Novikov:2020snp,Kotz:2025lio},
but, for the reasons discussed above, should be regarded as illustrative rather than a robust reconstruction.   
Despite these limitations, it is worth noting that the preferred values for $\beta$ are $\sim 1$, in support of some of the recent determinations~\cite{Barry:2021osv,JeffersonLabAngularMomentumJAM:2022aix,Novikov:2020snp}.

\textit{Summary and conclusion}: We have presented the first direct lattice QCD extraction of the flavor non-singlet pion PDF moment ratios $\braket{x^{n-1}}/\braket{x}$ for $n \in [3,6]$, using a novel method~\cite{Shindler:2023xpd} that employs the gradient flow as an intermediate regulator. 
Our results for the moment ratios show full agreement with phenomenological extractions. 
The precision achieved with a modest statistical sample is already comparable to phenomenology, opening the possibility of incorporating lattice QCD results from this method into global PDF fits~\cite{JeffersonLabAngularMomentumJAM:2022aix}, either directly in the form of ratios as presented here or as individual moments.  
In the latter case, determination of the normalization $\braket{x}$ is required; this can be obtained with standard lattice QCD techniques and renormalized in the $\overline{\MS}$ scheme either via the RI-MOM procedure~\cite{Martinelli:1994ty,Martinelli:1993dq} or by renormalizing the flowed fermion fields.  

This work demonstrates the viability of the method and provides meaningful results; a more detailed study of systematic effects will become necessary once higher statistical precision is achieved.  
Future investigations will include: 1) the impact of excited-state contamination,
2) a better understanding of cutoff effects and their possible reduction by exploiting different discretizations of the twist-2 operators based on the subduced hypercubic irreps, 
and 3) the impact of higher-order corrections in the perturbative matching. 
The PDF reconstruction from our results could also benefit from Bayesian approaches, such as Gaussian Processes~\cite{Dutrieux:2024rem,Candido:2024hjt}, which have the potential to stabilize the fits and provide less model-dependent reconstructions compared to the current parametric ansatz.

In addition to solving a long-standing challenge in the field, this method is complementary to other promising approaches~\cite{Aglietti:1998mz,Detmold:2005gg,Ji:2013dva,Monahan:2015lha,Radyushkin:2016hsy,Orginos:2017kos,Chambers:2017dov,Braun:2007wv,Ma:2014jla,Liang:2019frk}, and combining results from both should substantially enhance the impact of lattice QCD in this area.

Finally, this work opens the way for applying this method to other quantities of interest, including the proton PDFs, the flavor-singlet and gluon PDFs, and off-forward structure functions such as the Generalized Parton Distributions~\cite{Ji:1998pc,Radyushkin:1997ki}.  
\begin{acknowledgements}
\textit{Acknowledgements}: 
We thank Wally Melnitchouk for useful discussions on the reconstruction of PDFs from moments, and Martin Hoferichter for discussions on the chiral dependence of moments of PDFs. 
We thank the authors of Refs.~\cite{Gao:2022iex,Barry:2021osv} for providing their data. All phenomenological results in the comparison plots were obtained using the {\tt LHAPDF6} library~\cite{Buckley:2014ana}.
Numerical calculations for this work were performed using resources from the National Energy Research Scientific Computing Center (NERSC), a Department of Energy Office of Science User Facility, under NERSC award NP-ERCAP0027662; and the Gauss Centre for Supercomputing e.V. (www.gauss-centre.eu) on the GCS Supercomputer JUWELS~\cite{juelich2021juwels} at the Jülich Supercomputing Centre (JSC).
We also acknowledge the EuroHPC Joint Undertaking for awarding this project access to the EuroHPC supercomputer LEONARDO, hosted by CINECA (Italy) and LUMI at CSC (Finland).
The authors acknowledge support as well as computing and storage resources by GENCI on Adastra and Occigen (CINES), Jean-Zay (IDRIS) and Ir\`ene-Joliot-Curie (TGCC) under projects (2020-2024)-A0080511504 and (2020-2024)-A0080502271 as well as 2025-A0180516207.

A.F. acknowledges support by the National Science and Technology Council of Taiwan under grants 113-2112-M-A49-018-MY3 and 111-2112-M-A49-018-MY2. 
The work of R.K. is supported in part by the NSF Graduate Research Fellowship Program under Grant DGE-2146752. 
This research was supported by the National Research Foundation (NRF) funded
by the Korean government (MSIT)(No. RS-2025-02221606).
R.V.H. and J.T.K. acknowledge support by the Deutsche Forschungsgemeinschaft
(DFG, German Research Foundation)  under grants 460791904 and
396021762 - TRR 257 ``Particle
Physics Phenomenology after the Higgs Discovery''.
G.P. acknowledges funding by the Deutsche Forschungsgemeinschaft (DFG, German Research Foundation) - project number 460248186 (PUNCH4NFDI).
D.A.P is supported from the Office of Nuclear Physics, Department of Energy, under contract DE-SC0004658. 
A.S. acknowledges funding support from Deutsche Forschungsgemeinschaft (DFG, German Research Foundation) through grant 513989149 and under the National Science Foundation grant PHY-2209185. The work of A.W-L. was supported by the U.S. Department of Energy, Office of Science, Office of Nuclear Physics, under contract number DE-AC02-05CH11231.
The research of S.Z. is funded, in part, by l’Agence Nationale de la Recherche (ANR), project ANR-23-CE31-0019. For the purpose of open access, the author has applied a CC-BY public copyright licence to any Author Accepted Manuscript (AAM) version arising from this submission.
We acknowledge support from the DOE Topical Collaboration “Nuclear Theory for New Physics”, award No. DE-SC0023663. This manuscript was finalized at Aspen Center for Physics, which is supported by National Science Foundation grant PHY-2210452.

The {\tt OpenQCD}~\cite{openqcd}, {\tt Chroma}~\cite{Edwards:2004sx},  {\tt QUDA}~\cite{Clark:2009wm,Babich:2011np,Clark:2016rdz}, {\tt QDP-JIT}~\cite{6877336}, and {\tt LALIBE}~\cite{lalibe} software libraries were used in this work. 
Data analysis used {\tt NumPy}~\cite{harris2020array}, {\tt SciPy}~\cite{2020SciPy-NMeth}, {\tt gvar}~\cite{peter_lepage_2020_4290884}, and {\tt lsqfit}~\cite{peter_lepage_2020_4037174}.
Figures were produced using {\tt matplotlib}~\cite{Hunter:2007}.
\end{acknowledgements}

\bibliography{main}

\newpage
\newpage

\appendix

\section{End Matter}
\label{sec:EM}

\textit{Appendix A: Lattice QCD technical details}:
In this appendix we provide additional details on the lattice calculation underlying the results presented in the main text.

The main building block in the calculation of the ratio of moments presented in this work is the correlation function  
\bea
    &&C_n^{3\text{-pt}}( x_4  = \tau_s,\, y_4 = \tau_{\mcO}; t) =  \nonumber \\
    & & a^6 \sum_{\bx,\by} e^{i \bp \cdot (\bx - \by)} 
    \braket{P^{du}(\bx,x_4)\,\widehat{O}_n(\by,y_4;t)\,P^{ud}(0)}_c \,,
    \label{eq:3ptfull_mom} 
\eea
where the subscript $c$ indicates that only the fermionic connected contribution is included.  

In the non-singlet case there is no mixing with gluonic twist-2 operators, and the matrix elements for a spin-0 hadron are given by  
\be
\braket{h(p)|\widehat{O}_n|h(p)} 
= 2\,(p_{\mu_1}\cdots p_{\mu_n} - \text{traces})\, \braket{x^{\,n-1}} \,,
\label{eq:twist2ME}
\ee
with $h(p)$ a hadronic state of momentum $p$, and $\widehat{O}_n$ the trace-subtracted operator of Eq.~\eqref{eq:twist2}.  

In standard lattice QCD calculations, the renormalization and continuum limit of these operators is greatly complicated.  
For $n>2$, twist-2 operators mix with lower-dimensional operators, and this mixing is power-divergent in the lattice spacing $a$, preventing a controlled continuum limit.  
For $n=3,4$, specific irreducible representations (irreps) can be chosen to avoid such mixing~\cite{Martinelli:1987zd,Martinelli:1987bh}; however, only operators with all indices different transform irreducibly, which requires injecting spatial momentum in multiple directions in Eq.~\eqref{eq:twist2ME}. This leads to a severe deterioration of the signal-to-noise ratio, as confirmed in recent lattice studies~\cite{Alexandrou:2020gxs,Alexandrou:2021mmi}.  
For $n>4$, no irreps are available that are safe from power-divergent mixing, and consequently the extraction of higher Mellin moments of PDFs with local operators has remained out of reach.  

With the method adopted in this work, we are not constrained in the choice of Lorentz indices and can therefore select twist-2 operators with all temporal indices.  

In the evaluation of the correlation function in Eq.~\eqref{eq:3ptfull_mom}, we project to vanishing spatial momentum 
(Eq.~\eqref{eq:3ptfull}). For the calculation of the quark propagators we use a single randomly positioned, unsmeared $\mathbbm{Z}_4$ stochastic noise source per configuration, although for notational simplicity in Eq.~\eqref{eq:3ptfull} the source is written as if it were fixed at the origin.  

The three-point functions are measured at two sink-source separations per ensemble, using the sequential source-through-the-sink method~\cite{Bernard:1985ss},
and at several flow times up to $t/t_0 \simeq 2.7$, spaced in steps of $\sim 0.1$. 

Constructing the traceless operator bases for $n \in [2,6]$ requires $4$, $10$, $40$, $136$, and $544$ independent operators with unique Lorentz indices, which constitutes a major part of the computational cost.  
Moreover, each three-point function contains several terms arising from the discretized symmetrized covariant derivatives, with each term itself a sum of four contributions.  
To reduce this contraction cost, we employ several numerical improvements, detailed in Ref.~\cite{Francis:2024koo}.

If the flow-time footprint of $\widehat{O}_n(t)$ is sufficiently smaller than its Euclidean time separation from the pion creation and annihilation operators, i.e. $\sqrt{8t} \ll \tau_s - \tau_{\mcO}$ and $\sqrt{8t} \ll \tau_{\mcO}$, the three-point function of Eq.~\eqref{eq:3ptfull} admits the spectral decomposition  
\bea
\begin{split}
C_n^{3\text{-pt}}(\tau_s,\, \tau_{\mcO}; t) 
= & \sum_{k,k'} \frac{e^{-E_{k'} \tau_s}\, e^{-(E_k-E_{k'})\tau_{\mcO}}}{4 E_k E_{k'}}  \\
& \times Z^*_k Z_{k'} \braket{k'(\bzero)|\widehat{O}_n(t)|k(\bzero)} \,,
\end{split}
\label{eq:sprectral}
\eea
where $Z_k = \braket{0 | P^{du} | k(\bzero)}$, and the sum runs over all states $k$ in the spectrum.  
The ground-state contribution can then be isolated in the limit $\tau_s \gg \tau_{\mcO} \gg 0$.  
In this regime, for each flow time $t/a^2$, the ratio of three-point functions becomes independent of $\tau_{\mcO}$ (and of $\tau_s$), leading to the ratio of matrix elements $A_n(t)/A_2(t)$ given in Eq.~\eqref{eq:ratio_ground}.  
For certain values of $n$ on some ensembles, we observe enhanced correlated fluctuations in the ratio of Eq.~\eqref{eq:ratio_ground} as a function of $\tau_{\mcO}$, which complicates the identification of a clear plateau region. Fits including excited states give results consistent with simple plateau fits, indicating that possible systematic effects cannot be resolved with the present statistical precision. Excited-state contamination is treated conservatively in this work, as discussed in the "Lattice QCD calculation" section of the main text.
Future studies at lower pion masses and with higher statistics will require a more detailed analysis. 

\textit{Appendix B: Continuum limit}: 
In this appendix we summarize our continuum-extrapolation strategy and the treatment of cutoff effects for the correlation function entering our analysis. We first outline the Symanzik effective-theory expectations for the lattice artifacts relevant to our setup, then describe the specific simplifications that occur for ratios of flowed moments, and finally state the practical extrapolation procedure adopted in this work.

The cutoff effects of the correlation function in Eq.~\eqref{eq:3ptfull} can be analyzed within Symanzik effective theory~\cite{Symanzik:1983dc,Symanzik:1983gh,Luscher:1996sc}.  
For the exponential-clover (exp-cl) fermion action employed in this work~\cite{Francis:2019muy}, these effects can be discussed following the framework presented in Ref.~\cite{Luscher:2013cpa}. Once the lattice action is nonperturbatively O($a$) improved, as is the case here with the exp-cl action, the remaining O($a$) cutoff effects related to flowed operators are of order O($am$). There are potentially short-distance O($a$) effects that decay faster than exponentially with the Euclidean separations, $\tau_{\mcO}$ and $\tau_s-\tau_{\mcO}$, between the operator and the source/sink. We have checked numerically that the correlation function describing these effects is negligible in the range of $\tau_{\mcO}$ where we extract the matrix elements.

Besides further simplifying renormalization, the lattice calculation of ratios of flowed moments is also advantageous because O($am$) discretization effects cancel, as they depend only on the fermion content of the flowed operators~\cite{Luscher:2013cpa}. This implies cutoff effects that are parametrically of O($a^2$)~\cite{Shindler:2023xpd}. This situation is particularly favorable compared to the standard case, where the calculation of matrix elements of twist-2 operators requires a nonperturbative subtraction of O($a$) cutoff effects that depend on the moment $n$ under consideration.

\begin{figure}[!]
\centering
\includegraphics[width=0.5\textwidth]{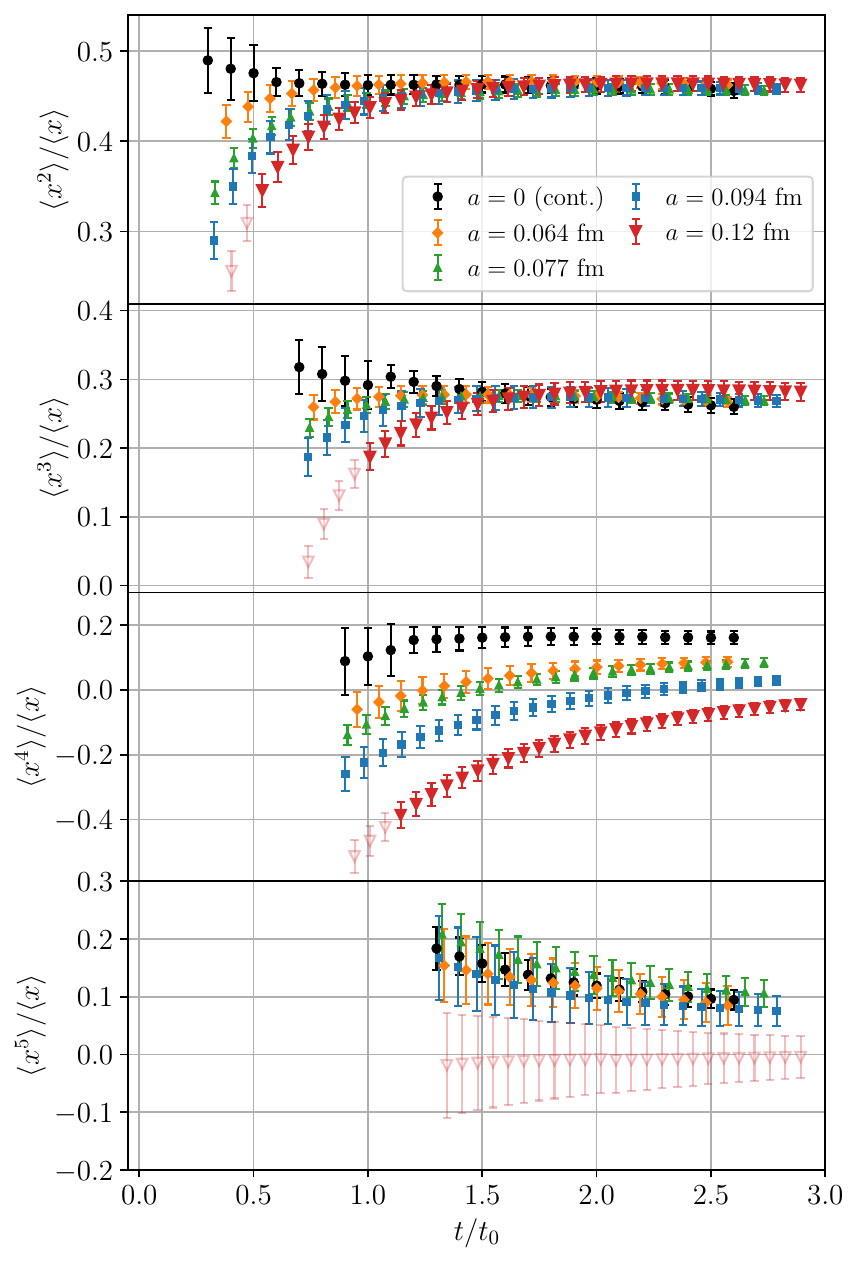}
\caption{Ratios of flowed Mellin moments $\langle x^{n-1} \rangle / \langle x \rangle$ for $n=3,4,5,6$ as functions of the flow time $t/t_0$.  
Results are shown for four lattice spacings $a \simeq [0.064, 0.077, 0.094, 0.12]$ fm and for their continuum extrapolation ($a=0$).  
Points in lighter colors are excluded from the continuum limit (see main text).}
\label{fig:allfinal}
\end{figure}

In Fig.~\ref{fig:allfinal} we show the flow-time dependence of the ratios of flowed moments, 
$r_n = \braket{x^{n-1}}/\braket{x}$, at each lattice spacing together with the resulting continuum limit (black points).  
The O($a^2$) discretization errors for these ratios can be parametrized as  
\be
r_n(a,t) = r_n(t) + \delta_2(t)\,a^2 + \cdots \, ,
\label{eq:cont_a2_form}
\ee
where $\delta_2(t)$ includes contributions that are power-enhanced, i.e. proportional to $a^2/t$.  
The observed flow-time dependence of $r_n$ in Fig.~\ref{fig:allfinal} confirms this expectation.  

For the purpose of this Letter, we adopt a simple procedure in which each moment is extrapolated to the continuum limit independently at every flow time.  
The extrapolation uses the set of lattice spacings closest to the continuum that are well described by Eq.~\eqref{eq:cont_a2_form}, with a trend toward including more lattice spacings at larger flow times.  
The range in $t/t_0$ for which this approach is valid depends on both the moment and the flow time, and defines the $\left(t/t_0\right)_{\text{min}}$ used in the subsequent extrapolation to vanishing flow time.  
The results presented here are consistent, in both central values and uncertainties, with a more sophisticated analysis that more thoroughly assesses the systematic uncertainty associated with the continuum extrapolation.

\clearpage

\onecolumngrid
\setcounter{page}{1} 

\section{Chiral Corrections to Ratios of Pion PDF Moments}

\begin{figure}[h]
    \includegraphics[width=0.5\linewidth]{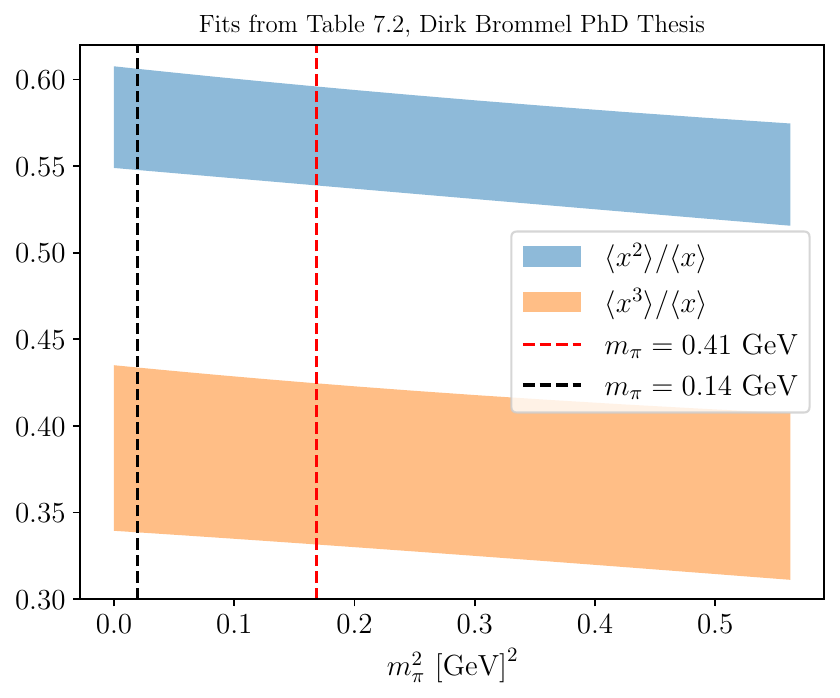}
    \caption{Pion mass dependence of the ratios $\langle x^2\rangle/\langle x\rangle$ and $\langle x^3\rangle/\langle x\rangle$ reconstructed from the non-singlet PDF moments $\langle x\rangle$, $\langle x^2\rangle$, and $\langle x^3\rangle$ reported in Ref.~\cite{Brommel:2007zz} (Table 7.2).  
Covariances were not available in Ref.~\cite{Brommel:2007zz} and have therefore been neglected in forming these ratios.  
The vertical dashed lines indicate the pion masses $m_\pi = 0.41$~GeV (red) and $m_\pi = 0.14$~GeV (black).}
\label{fig:chiral}
\end{figure}

The chiral corrections to twist-2 matrix elements were derived in Ref.~\cite{Arndt:2001ye}. Lattice calculations of the connected contributions to the pion unpolarized PDF moments with $n=2,3,4$ were presented in Refs.~\cite{Brommel:2005ee,Brommel:2007zz}, using irreps without mixing and pion masses in the range $0.3$-$1.2~\text{GeV}$. Based on the chiral fits reported in Table~7.2 of Ref.~\cite{Brommel:2007zz}, Fig.~\ref{fig:chiral} shows the expected pion-mass dependence of the ratios $\braket{x^2}/\braket{x}$ and $\braket{x^3}/\braket{x}$. These results are not yet renormalized to $\overline{\text{MS}}$ at $\mu=2~\text{GeV}$, so the values at the physical point differ from the final results of Ref.~\cite{Brommel:2007zz}.
These results suggest that the chiral dependence of the ratios is small and subdominant compared to our current uncertainties, as seen by noting the shift between the pion mass used in this work (red dashed line) and the physical one (black dashed line). More data near the physical point are nevertheless required to confirm this. In Refs.~\cite{Brommel:2005ee,Brommel:2007zz} the lightest pion mass was still relatively heavy at $m_{\pi}\simeq300~\text{MeV}$, and the leading logarithmic terms may become more relevant closer to the physical pion mass. The chiral behavior of $\braket{x^{n-1}}$ in the pion has also been investigated in quenched studies~\cite{Detmold:2003tm,Guagnelli:2004ga,Capitani:2005jp}.

\end{document}